# Developing arrayed waveguide grating spectrographs for multi-object astronomical spectroscopy

Nick Cvetojevic,[1,2,4] Nemanja Jovanovic,[1,2,3] Jon Lawrence,[1,2,3] Michael Withford,[1,2,4] and Joss Bland-Hawthorn[5,6]

[1]*MQ Photonics Research Centre, Department of Physics and Astronomy, Macquarie University, NSW 2109, Australia*
[2]*Centre for Astronomy, Astrophysics and Astrophotonics, Macquarie University, Australia*
[3]*Anglo-Australian Observatory, PO Box 296, Epping, NSW 2121, Australia*
[4]*Centre for Ultrahigh Bandwidth Devices for Optical Systems*
[5]*Sydney Institute for Astronomy, School of Physics, University of Sydney, NSW 2006, Australia*
[6]*Institute of Photonics and Optical Science, School of Physics, University of Sydney, NSW 2006, Australia*

*Author e-mail: nick.cvetojevic@mq.edu.au*

**Abstract:** With the aim of utilizing arrayed waveguide gratings for multi-object spectroscopy in the field of astronomy, we outline several ways in which standard telecommunications grade chips should be modified. In particular, by removing the parabolic-horn taper or multimode interference coupler, and injecting with an optical fiber directly, the resolving power was increased threefold from $2400 \pm 200$ (spectral resolution of $0.63 \pm 0.2$ nm) to $7000 \pm 700$ ($0.22 \pm 0.02$ nm) while attaining a throughput of $77 \pm 5\%$. More importantly, the removal of the taper enabled simultaneous off-axis injection from multiple fibers, significantly increasing the number of spectra that can be obtained at once (i.e. the observing efficiency). Here we report that ~ 12 fibers can be injected simultaneously within the free spectral range of our device, with a 20% reduction in resolving power for fibers placed at 0.8 mm off-centre.

**1. Introduction**

Multi-object spectroscopic instruments have the capacity to measure the spectra of multiple (often hundreds) of stellar or extragalactic sources simultaneously. This powerful technique allows astronomers to make observations many orders of magnitude faster than observing a single object at a time, allowing for large scale surveys of the universe to be carried out [1]. For some multi-object spectrographs, optical fibers are accurately positioned at the focal plane of the telescope, which transport the light to a spectrograph. The next generation of major ground-based optical and near-infrared telescopes, such as the Giant Magellan Telescope (GMT) or the European Extremely Large Telescope (E-ELT), will have aperture sizes expected to exceed 20 m in diameter, far greater than existing telescopes. This has a major impact on the spectrographs implemented on the telescope, as for seeing-limited observations the size of the instrumentation grows in proportion to the aperture ($D$), and the cost of the instrument increases as $D^2$ or faster [2,3]. This poses fundamental problems for practical implementation of ELT instruments in terms of flexure and thermal stability of multi-object spectrographs.

The use of integrated photonic circuits to create a new generation of miniature spectrographs for astronomy, with existing technologies already developed by the telecommunications industry, was first suggested in 1995 by Watson [4,5], and treated more comprehensively by Bland-Hawthorn & Horton [2]. One of the concepts was to use a "spectrograph on a chip" dubbed the integrated photonic spectrograph (IPS), which would potentially be orders of magnitude smaller and less expensive than existing astronomical spectrographs while offering superior thermal stability and being less prone to flexure [2,6]. This photonic approach to spectrograph design allows for small, mass-fabricated, modular components to be used in favor of large, custom-built components used in existing spectrographs.

The IPS outlined in [2], composes a single-mode optical fiber, which directly feeds the input signal into a silica-on-silicon array waveguide grating (AWG) chip. A typical AWG chip is depicted in Fig. 1 below. It uses input single-mode waveguides that feed into an input multiplexor, or free-propagation zone (FPZ). The FPZ is effectively a slab waveguide with a step index which allows the light to diverge in the plane of the chip onto a parallel array of closely spaced single-mode waveguides at the far end. These in turn feed a demultiplexor (another FPZ). As each waveguide in the array is incrementally longer than the previous one by a constant amount, the array of waveguides behave analogously to the teeth of a blazed grating and create multiple, phase-shifted point-sources of light. The light then interferes in the second FPZ such that a dispersed spectrum is formed at the output end of the FPZ. By designing the path length increment between waveguides in the array, the position at which a specific wavelength is focused on the end-face of the slab can be tailored. For telecommunication applications, the dispersed light at the output of the second FPZ is typically coupled into a series of waveguides [7]. Discrete output channels are not useful for astronomical applications; rather a continuous spectrum is preferred. Hence the output waveguides are removed entirely with the entire end-face of the chip being imaged using a pixel based imaging array. Our polishing creates a flat output surface, while the input and output ends of the FPZs are in fact curved, and indeed match the so-called Rowland curvature[7]. The Rowland curvature is tailored in such a way that the waveguides of the array collect the diverging light in-phase (i.e. along a given wavefront), for example. This means that the flat polished surface was not ideal but was enough to demonstrate the concepts outlined in this paper. The penalty for not perfectly matching to the Rowland curvature is elucidated further on.

We have previously demonstrated the feasibility of using AWG chips as spectrographs for astronomy with a prototype IPS used to observe the OH atmospheric emission lines without a telescope. The observations were made in the astronomical H-band atmospheric window (1485 – 1825 nm) that encompasses the telecommunications C-band (1520-1570 nm) for which the chips were optimized [8]. However, to create an IPS that is practical for astronomical multi-object spectroscopy, the capacity of taking multiple input signals simultaneously, and adequately imaging them, must be characterized. The requirement for multiple input signals is driven by the need to convert from a multi-mode format provided at the telescope focal plane to a single mode format required by the AWG. Further, the single-mode nature of the AWG injection necessitates a more complex telescope interface for the chips and is discussed in more detail in section 3.

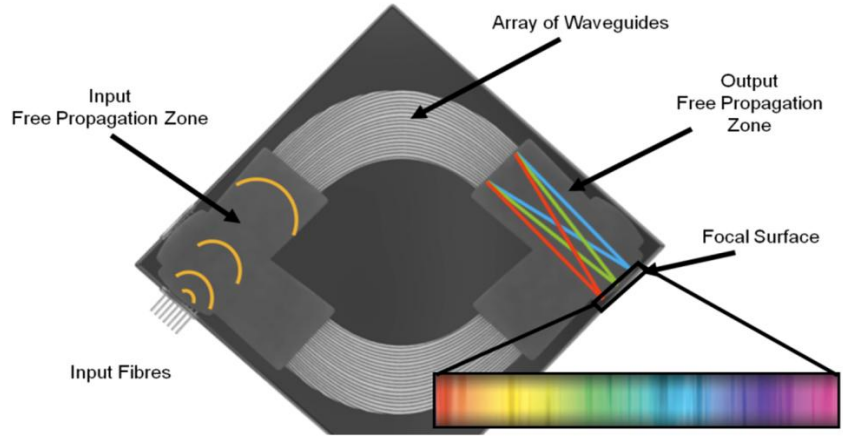

Figure 1. Illustration of an AWG-chip configuration that has been modified for astronomical use by removing the output waveguides. The input fibers/waveguides to the far-left couple the light into the input FPZ, which, after dispersing the light, couples it into the array of closely spaced waveguides, and finally interfering it in the output FPZ forming a spectrum on the end-face of the chip we dub the 'focal surface'. Commercially available AWG-chips typically have two AWG structures on the one chip so as to maximize the real estate of the substrate, as illustrated above. In our experiments, we use only one of the AWG structures per chip; however it is important to note that one could potentially use both structures simultaneously.

Astronomical spectrographs have substantially different requirements and tolerances than that of the typical applications of AWGs as wavelength division multiplexers and de-multiplexers (WDMs) for telecommunication. In this paper we present results on the impact conventional AWG chip designs, such as parabolic-horn couplers, and multimode interference couplers (MMIs), have on spectroscopic applications of the chips, relating to specific parameters important for astronomy. Furthermore, we explore the practicality of directly coupling single-mode optical fibers to the input FPZ, which we are calling the direct-fiber launch method. Lastly and most importantly we show results demonstrating the feasibility of directly launching, and taking the spectra of multiple fibers simultaneously, which leads to an increased astronomical observing efficiency.

## 2. Parabolic-horn tapers and MMIs' impact on AWG-spectrograph performance

The AWG can be thought of as an imaging spectrograph, where the shape of the input intensity distribution on the front of the first FPZ is re-imaged onto the end of the second FPZ [9-11]. Unmodified AWG-WDMs focus the spectrally dispersed light into separate waveguides at the output of the demultiplexor, thus sampling a continuum into discrete wavelength channels. If single-mode waveguides are used for injection and collection of the light, any misalignment between the position of the collecting waveguide and the re-imaged spot will lead to a mode mismatch between the two, and result in a loss in the system. Misalignments can arise from errors in fabrication or temperature changes, but even if the chip itself is stabilized, wavelength instabilities in laser-sources used in telecommunications will cause a mismatch. To alleviate these restrictions typically a parabolic-horn taper, or MMI coupler, is employed where the input waveguides meet the input FPZ (see Fig 2(a)), and mirrored for the output waveguides [12-14]. By careful control of both the width and length of the tapers, interference between the fundamental and $2^{nd}$ order modes broadens the intensity profile yielding a double-peaked electric field distribution which is typically two or three times wider than that of a standard single mode input waveguide (Fig 2(c)) [13]. This broader intensity profile is less susceptible to misalignment with the sampling waveguides at the output of the device. However, when utilizing AWGs as spectrographs all output waveguides are removed, thus creating a flat output plane from which a continuous spectrum can be obtained. Hence parabolic tapering of input waveguides for loss minimization is redundant, and in fact, as we will explain, is disadvantageous for astronomical applications.

The AWGs we use for our experiments have identical parameters; their central operating wavelength was 1540 nm at the $m = 27$ diffraction order, ~22 mm focal length FPZs, and the waveguide array consisted of 428 waveguides with a length increment of ~28 μm. Before the output waveguides were removed, the device was designed to operate with a 100 GHz (0.8 nm) output channel separation, and 40 output waveguide channels. The waveguides were removed by polishing the chip up to the input FPZ by the same method used to remove the output waveguides. As there was no obvious difference in the overall throughput between the polished and unpolished chips, we conclude that the there was no significant surface scattering and hence the surface roughness was of optical quality. Two different AWG chips were characterized to explore the effects of parabolic tapers, the first containing parabolic-tapered input waveguides on the chip, and the second having all input waveguides and tapers removed, shown in Fig. 2(a) and (b) respectively.

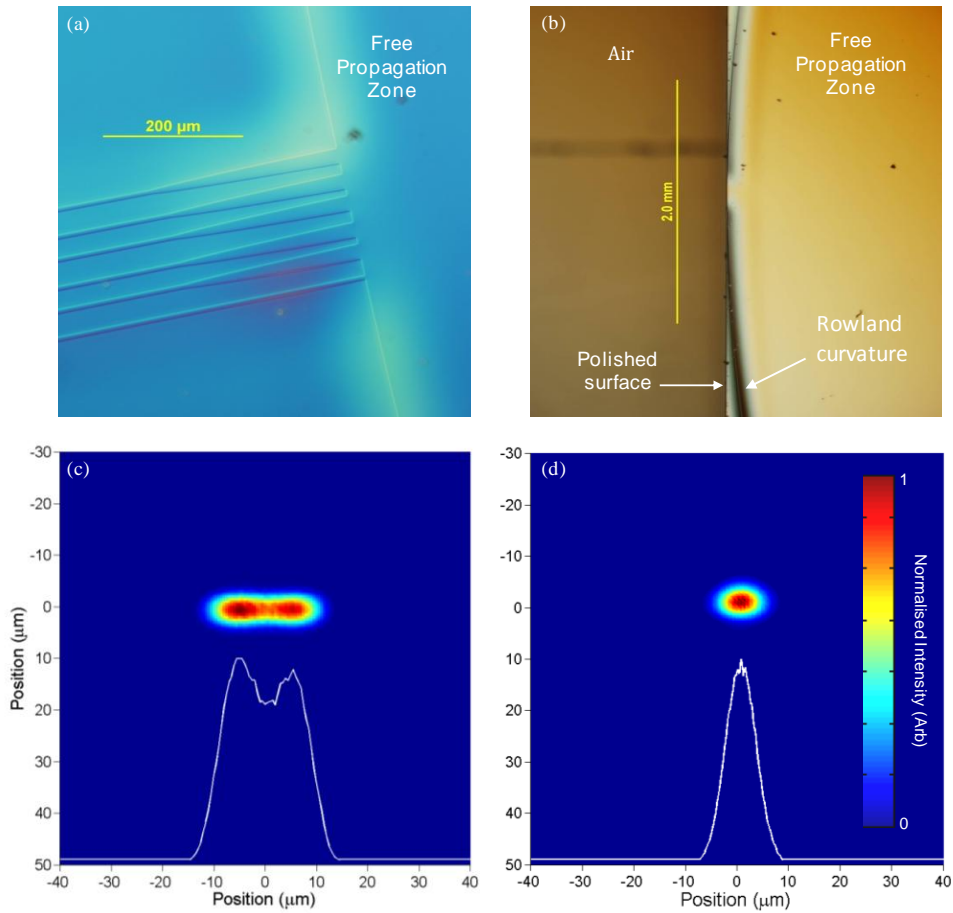

Figure 2. (a) The parabolic-type tapers imaged using differential interference contrast microscopy at the input of a free propagation zone of an unmodified AWG chip. (b) The input face of the chip after polishing. The fibers are directly butt-coupled up to the free propagation zone of the chip. Comparison of the point-spread-function from the parabolic input waveguide taper and that of the direct fiber launch can be seen in (c) and (d) respectively. The intensity for both profiles is linear in arbitrary units.

The chips were probed using the setup in Fig. 3. A tunable laser diode operating in the C-band was coupled into the two chips via a single-mode fiber, while a flux-calibrated InGaAs near infra-red array detector was used to image the field profiles emanating from the chips. This allowed accurate measurements of both the throughput efficiency and the shape of the point-spread-function, which in turn were used to determine the resolution of the AWG spectrograph. For the chip with no input waveguides we used a direct-fiber launching technique, where a commercial single-mode optical fiber (SMF-28) was butt-coupled to the input face of the chip, launching the light directly into the input FPZ. The coupling was maximized and the fiber accurately aligned using a 5-axis flexure translation stage with a 0.5 μm positional accuracy. In the case of the chip with the lead-in waveguide still intact, the light was launched into the waveguide by butt-coupling to the SMF-28 fiber. Fresnel reflections were minimized with the use of index-matching gels.

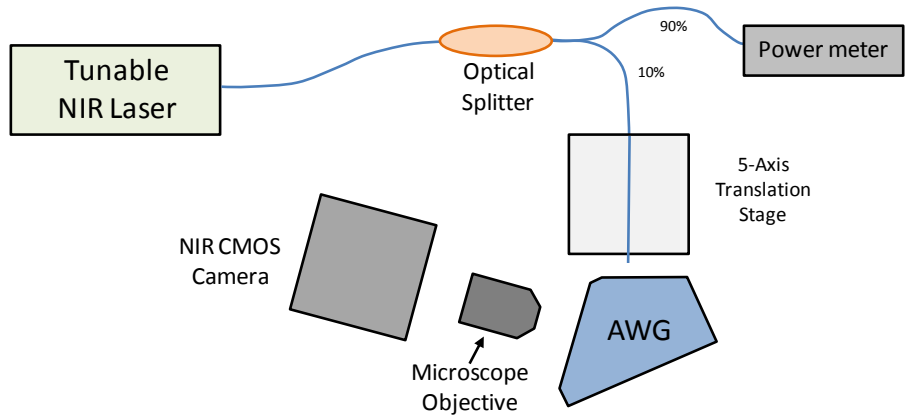

Figure 3. Experimental setup for the concurrent characterization of the chips' point spread function and throughput efficiency.

The output mode profile (which is now analogous to the spectrograph's output point-spread-function) of the AWG with the taper in contact showed the expected, distinctly broadened, double-peaked profile, with a physical FWHM of 18.7 ± 0.7 μm and an associated spectral FWHM of 0.63 ± 0.2 nm shown in Fig. 2(c). This translates to a spectrograph resolving power $R = \lambda/\Delta\lambda$ of ~ 2400. By contrast, the direct-fiber launched AWG chip demonstrated a Gaussian profile (Fig. 2(d)) with a physical FWHM of 6.6 ± 0.6 μm, and a spectral FWHM of 0.22 ± 0.02 nm, yielding $R$ ~ 7000. It should be made clear that this is solely a result of the smaller physical size of the injection PSF from the single-mode fiber as compared to the horn-taper rather than the specific shape of the PSF injected. The resulting resolving power is much closer to the theoretical maximum of a diffraction-limited device where for $N$ combining beams, the limiting resolving power is $R_T = m\,N$ ~ 11,600 [2,3]. The discrepancy between the measured and theoretical $R$ is attributed to both manufacturing imperfections as well as the fact that the waveguide array was illuminated with a Gaussian intensity distribution (flat-top would be ideal). The throughput was determined by comparing the injected power, measured at one port of a 90:10 splitter with a power meter, to the total encircled power as measured by the flux-calibrated camera, shown in figure 3. The two measurements were taken simultaneously to minimize any fluctuations in the power of the laser. The peak efficiency was identical for both chip configurations to within our experimental error and was measured to be ~ 75 ± 5%. As the throughput did not change when the input tapers were removed then the direct-fiber launch is suitable for use. Indeed it is preferable as there is a three-fold increase in the resolving power associated with the removal of the input taper. Finally, by redesigning the AWG's it has been demonstrated that the size and shape of the injection PSF can be tailored to suit a particular application [15], however optical fibers offer the freedom of positioning multiple injection ports at arbitrary positions along the input.

## 3. Direct multiple fiber input

Fiber-fed astronomical instruments use multimode fibers (typically 50-150 μm cores) to ensure that the focused light from the telescope is adequately captured and transported down the fiber. This is done as it is difficult to efficiently focus atmospherically perturbed wavefronts into single-mode fibers, even with the aid of adaptive optics [16,17]. Such large diameter cores result in a highly multimode signal being injected into the spectrographs. These transport fibers are not compatible with AWG technology as any introduction of modes higher than the fundamental mode severely degrades the performance of the devices. Hence conversion between multimode and single-mode formats is necessary. Unfortunately, due to the brightness theorem, focusing multiple modes into one single mode is impossible. The best option to increase both observational and throughput efficiency is to separate the modes in a multimode fiber into a number of individual single modes using devices such as the photonic lantern [18-21]. The lantern is comprised of an array of isolated SMF cores which are adiabatically tapered into one MMF core such that the supermodes of the SMF array evolve into the modes of the MMF core, and vice versa. Hence, multimode light captured by the telescope can be efficiently separated into multiple SMFs, enabling devices that only operate in the single mode regime, such as fiber Bragg gratings used for OH emission line suppression [22] and the AWG based spectrograph.

As an example, under seeing-limited conditions, a typical telescope with a 4 m primary mirror diameter, will have a PSF which can support hundreds of modes, scaling up to the ELTs, with mirror sizes greater than 25 m, we potentially face thousands of modes for a single source. Coupled with the fact that for multi-object spectroscopy we require simultaneous observations of multiple sources, this will require thousands of SMFs to interface with our AWG spectrographs. If each AWG chip only takes one SMF input (like the first prototype [8]), thousands of AWG chips would be required for a complete instrument, significantly increasing its total cost and volume and reducing the inherent benefits in terms of stability and robustness. It is therefore desirable to increase the number of SMFs used to inject into the AWGs.

By using the direct-fiber launch technique it is possible for an almost arbitrary positioning of the injection fiber along the input free propagation zone, and moreover it enables the possibility to interface multiple fibers with the chip simultaneously. We explored the feasibility of launching multiple single-mode fibers by examining how the throughput efficiency and PSF behave with relation to the input fiber position.

The experiments in 3.1 were conducted by translating one SMF along the input surface at 100 μm increments, while simultaneously imaging the output and calculating the throughput efficiency, using the same methodology described in section 2. The throughput efficiency at the central operating wavelength of 1540 nm was measured to be 77 ± 5% for fibers placed at the centre of the FPZ, gradually decreasing as a function of off-centre fiber position (Fig. 4(d)).

*3.1 Point-Spread Function aberrations from off-axis launch*

As shown in Fig. 2(d), the PSF has a Gaussian profile for fibers that are launched at the centre of the free propagation zone (also shown in Fig. 4(b)), with a resolving power of $R \sim 7000$ (see Fig. 4(d)). However, by injecting fibers with an increasing offset from the centre of the FPZ, the PSF becomes increasingly broadened and aberrated as shown in Fig. 4 (a) & (c). The broadening of the PSF reduces the $R$, and indeed, at +/- 0.8 mm the resolving power of the spectrograph is ~ 20 % lower.

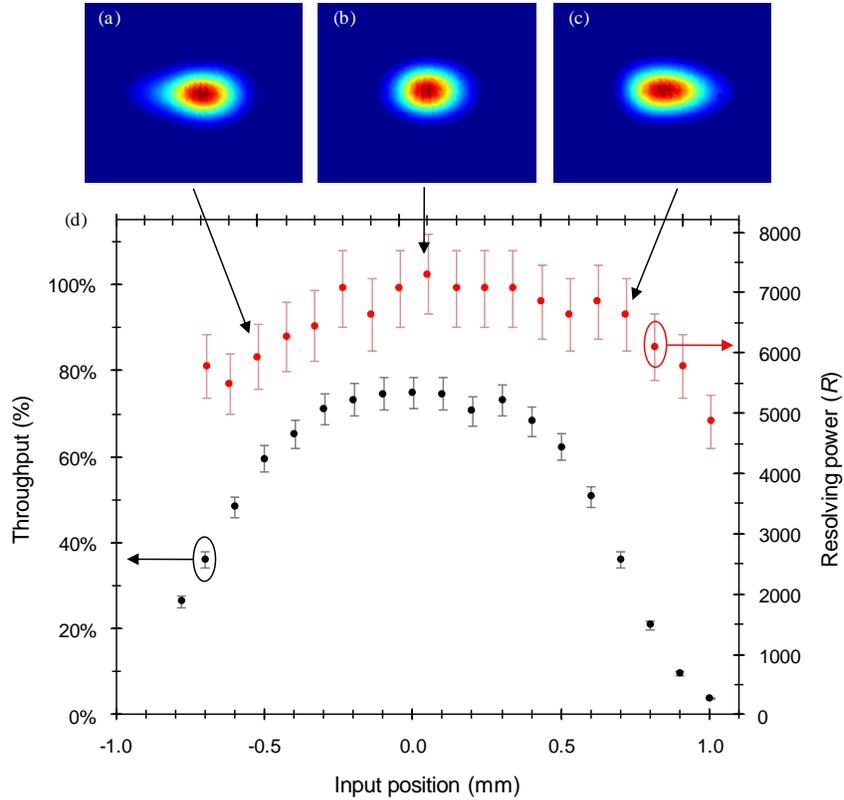

Figure 4. The resolving power is shown for input fiber positions across the input surface (red). The resolving power drops off for fibers placed further off centre. The throughput as a function of input fiber position is shown in black. The measurements were taken at 1540 nm.

The angle at which the launch fiber interfaces with the chip was kept consistently perpendicular to the polished facet. As such it was not aligned with the Rowland curvature of the input FPZ (which is depicted in Fig. 2(b)). Therefore, as we moved further from an on-axis launch the angle that the light was injected into the input FPZ increased monotonically from normal incidence. This resulted in a mismatch between the wavefronts and the array of waveguides at the far end of the input FPZ, which we believe caused the PSF to skew at the output of the device, in the case of the off-axis launch as seen in Fig. 4(c). The increasingly skewed nature of the output PSF as a function of increased offset of the injection fiber from the centre was corroborated by using an AWG model constructed in the beam propagation tool 'BeamProp'.

The aberration can in principle be corrected for by matching the input fiber launch angle to the Rowland curvature of the input FPZ, and imaging along the Rowland curvature of the output FPZ. Several techniques for achieving this, including one that involves the redesign of the geometry of the FPZ of the AWG chip, have previously been demonstrated [23,24]. However, performance optimization of the device presented here was beyond the scope of this work. Nonetheless, it is still possible to use the current device by first assigning a cutoff for acceptable resolving power and throughput, which in turn will determine the number of fibers which can be used to inject into the AWG chip for that required level of performance.

*3.2 Central wavelength shift due to off-axis launch*

For astronomical observations the polished end-face of the AWG chip's output FPZ is imaged using a 2D InGaAs array to entirely sample the continuous spectra across the whole free-spectral range. We therefore characterized the efficiency profile for the 27th order by scanning across the full wavelength range of our input laser (between 1500 and 1580 nm in 5 nm intervals), shown in Fig. 5. Multiple fibers were placed at the input using a linear fiber array with a 127 μm spacing between neighboring cores. This spacing is the minimum allowable as a result of the physical outer diameter of SMF being 125 μm. While the array comprised 16 fibers, the results presented in Fig. 5 are only for a subset of five of the fibers. The first fiber was positioned at the center with the others positioned at 127 μm increments increasingly further off-centre in one direction.

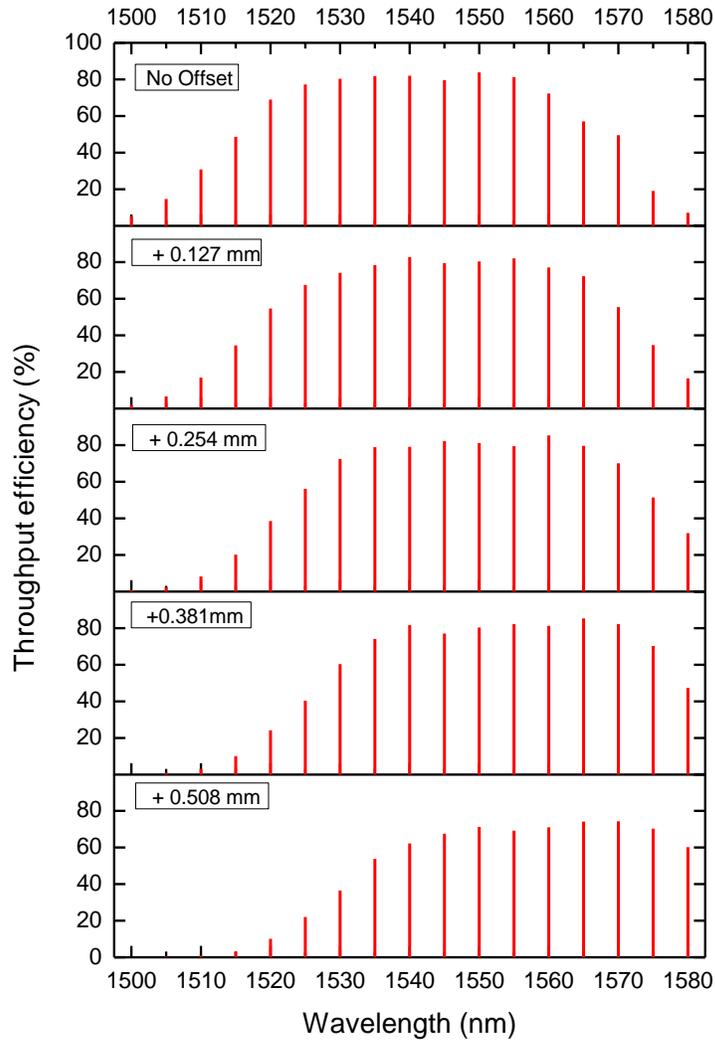

Figure 5. Efficiency across an 80 nm bandwidth for the 27th order measured in 5 nm wavelength intervals. The experimental uncertainty for all efficiency measurements presented in this plot is 5%. The fiber positioned at the centre of the FPZ is shown at the top, with progressively more position offset for subsequent fibers. The central wavelength for the order clearly shifts to longer wavelengths proportionally to off-axis fiber translation.

The efficiency curve, while maintaining the same shape, is different for each of the input fibers. The central wavelength of each order shifts to longer wavelengths with respect to the off-axis position of the input fibers. However, the profile of the efficiency curve is consistent across the physical output end-face of the chip for all input fibers, as they are shifted. The impact this has on practically implementing an AWG-based spectrograph with multiple inputs is that the response for each fiber will be different based on its input position, and hence needs to be characterized before implementation. Furthermore, despite the slight off-axis injection there is still significant overlap with the on-axis output spectrum, hence cross-dispersion in the orthogonal direction is required to obtain a meaningful measurement. The shift of the central wavelength as a function of the input fiber position imposes a limit to the number of fibers that can be used in total. This is because for a sufficiently large offset in injection position from the center, the peak wavelength of the AWG free-spectral range overlaps with the neighboring order of the on-axis fiber, which is spectrally indistinguishable. In this case it would not be possible to separate these two spectra with cross-dispersion. For our device, with a free-spectral range of ~60 nm, a fiber spacing of 127 μm, this would occur at the $13^{th}$ off-axis fiber. Hence a maximum of 12 fibers would ensure straightforward distinguishability between neighboring orders.

## 4. Conclusion

In our previous work [8], we demonstrated that AWG chips have the potential to be used as astronomical spectrographs, with a successful detection of atmospheric emission lines. However, such a device (one single-mode input per chip) is not practical for observing large numbers of astronomical targets, or when interfacing with most telescopes. By characterizing the direct launch of multiple fibers simultaneously we have shown that it is possible to increase the number of spectra detected by an AWG-based spectrograph. This can increase observational efficiency in the case of a telescope aided by adaptive optics (where light is directly coupled into SMFs) by allowing multiple sources to be observed per chip. For the seeing limited case, which requires a photonic lantern device to couple into SMFs, we are able to use one chip per source, which is still more efficient than the tens of chips that would be required if a single SMF is injected per chip. The limitation is that this technique requires cross-dispersion optics and a 2D detector, which adds extra complexity to the system. Alternatively, one can conceive of forgoing multiple fiber injection in favor of using cheaper 1D detector arrays, however this will require many more AWG chips to be used. These two approaches would need to be balanced to compare their respective economic viability, as at the moment it is unclear. The conventional practice of using parabolic-horn tapers or MMIs for increased throughput efficiency was shown to be redundant and detrimental for astronomical applications. By removing the tapers and using a direct-fiber launch we increased the resolving power from 2400±200 to 7000±700 while maintaining the same efficiency of 77 ± 5%. Simultaneous off-axis launch of multiple fibers was characterized, with fibers placed up to 0.8 mm off-centre showing a 20 % decrease in resolving power due to mismatch between the fiber launch angle and the Rowland curvature of the FPZ. Further we found that the efficiency curve for a given order will change depending on fiber position, with the central wavelength being shifted by ~20 nm for every 500 μm translation of input position.

We have shown that commercially available AWG chips can be modified to simultaneously image spectra that come from multiple input fibers. This can potentially increase the observational efficiency of an AWG based spectrograph for astronomy, and indeed other applications where such spectrographs can be used. With the ability to image multiple spectra one can now have control fibers to monitor background noise, as well as greatly increasing the number of sources that can be monitored per chip, potentially reducing the number of chips that need to be used, and hence the cost.

Finally, other AWG parameters, such as the FPZ length, the number of waveguides in the array, the operating wavelength and spectral order, and so forth, can be modified to further facilitate its use as an imaging spectrograph designed specifically for astronomy. We will also investigate their use in spectropolarimetry and their performance at cryogenic temperatures. These are some of the issues to be addressed in future work.

**Acknowledgements**

This work was produced with the assistance of the Australian Research Council (ARC) under the ARC Centres of Excellence and the Australian Astronomical Observatory. Elements of this work were supported by the EU via the Opticon Integrated Infrastructure Initiative of Framework Programme 7. Nick Cvetojevic acknowledges support through the MQRES grant scheme from Macquarie University. Joss Bland-Hawthorn is supported by a Federation Fellowship from the ARC.